# Growth of SiO$_2$ microparticles by using modified Stöber method: Effect of ammonia solution concentration and TEOS concentration

Shrestha Bhattacharya[1,2], Aishik Basu Malick[1,2], Mrinal Dutta[2,a)], Sanjay K. Srivastava[2,3], P. Prathap[2,3] and C.M.S. Rauthan[2]

[1]*Department of Energy Engineering, Central University of Jharkhand, Jharkhand-835205, India*
[2]*PV Metrology Group, Advanced Materials Devices and Metrology Division, CSIR-National Physical Laboratory (NPL), New Delhi-110012, India*
[3]*Academy of Scientific and Innovative Research, CSIR-NPL Campus, New Delhi-110012, India.*

a)*Corresponding author: sspmd.iacs@gmail.com*

**Abstract.** The unique structural features and suitability of the SiO$_2$ micro particles in different application areas have mobilized a worldwide interest in the last few decades. In this report a classical method known as the Stöber method has been used to synthesize silica microspheres. These microparticles have been synthesized by the reaction of tetraethyl orthosilicate (Si(OC$_2$H$_5$)$_4$, TEOS) (silica precursor) with water in an alcoholic medium (e.g. ethanol) in the presence of KCl electrolyte and ammonia as a catalyst. It has been observed that the size of the microparticles closely depends on the amount of the TEOS and ammonia. A decrease in the size of micro particles from 2.1μm to 1.7μm has been confirmed as the amount of TEOS increases from 3.5 ml to 6.4 ml respectively. In similar way a decrease in the diameter of the micro particles from 2.1 μm to 1.7 μm has been observed with increase in the ammonia content from 3 ml to 9 ml.

## INTRODUCTION

The unique structural features of the silica micro particles have attracted a large attention since the last few decades. It is now being used for a large number of applications which include lithium ion batteries, catalysis, drug delivery, anti-reflective coating materials and also for cosmetics [1-5]. Due to mobility and high mechanical strength [6] SiO$_2$ particles are also used for column packing, structural ceramics material ink additives, etc. Beside these applications SiO$_2$ micro particles are also used to fabricate microwires solar cell using nanosphere lithography technique. In order to get different sizes of the silica particles, various methods of preparations have been adopted. Among the various methods used to prepare silica microspheres some are mechanical alloying method, micro emulsion method, hydrothermal synthesis method, precipitation, sol gel method and radiation synthesis method etc. [7-9]. The physical as well as the optical properties of the silica particles depend on the size of the particles.

Mostly the solution gelation commonly known as sol gel method is preferred over all other methods as this method involves simple chemistry and involves low cost techniques. In the sol-gel method a classical method known as the Stöber method is used to produce silica spheres [10, 11].

In this process silica precursor tetraethyl orthosilicate (Si(OC$_2$H$_5$)$_4$, TEOS) is first reacted with water in an alcoholic medium typically ethanol in the presence of ammonia as a catalyst [12, 13] and also KCl electrolyte. It has been found that adding KCl could effectively increase the size of the silica particles.

Recently Lei et al. [14] reported that using Stöber method silica micro particles having a diameter of about 1-3 μm could be obtained. He tried two different experiments. In one the two solutions were allowed to react for 15 hours and after centrifugation and washing with ethanol he found that the diameter of the micro particles was about 1μm. Further they continued the reaction in which a solution containing ethanol and TEOS was injected for 6 hours at a rate 0.1-0.2ml/min in a solution containing KCl, ethanol, water and ammonia and further the reaction was left for 5 hours. In this they synthesized micro particles of 2-3μm size.

In this work synthesis of SiO$_2$ particles up to 2.1 μm using the Stöber process has been reported. Variation of the size of the microparticles with the variation in the amounts of TEOS and ammonia has been investigated.

## EXPERIMENTAL

### Materials

TEOS was purchased from Sigma Aldrich. KCl and the ethanol were purchased from the Merck life Science Private Limited. The ammonia solution (28-30%) was also purchased from Merck life Science Private Limited. All the other chemicals and the reagents were used as they were received without any sort of purification.

### Preparation of the SiO$_2$ particles

SiO$_2$ micro particles were synthesized using simple modified Stöber method in an ethanol solution along with ammonia solution which is used as a catalyst. The reaction is carried out at room temperature in a 250 ml beaker with mechanical stirring of about 200 rpm. Two solutions were made. Solution A was a mixture of KCl (15mg), ethanol (60ml), Water (7ml) and ammonia solution. Ammonia solution was varied amount wise as 1.5 ml, 3 ml, 6 ml and 9 ml respectively. Solution B was a mixture of TEOS (3.5ml, 6.4ml) and Ethanol (38.5ml, 38.6ml). The solution B was supplied to solution A using a syringe pump. After the injection and further reaction, the micro particles were collected in a centrifuge tube and purified by centrifugation and washing by ethanol three times. Finally, the micro particles were dried either by natural evaporation or by evaporating the ethanol using hotplate at about 80 $^0$C. The micro particles obtained in this way are then dried and dispersed in DMF. They are then spin coated on the cleaned Si substrates.

### Characterization

The size and the distribution of the silica micro particles synthesized are then viewed by an optical microscope. However high resolution images of the microspheres were taken using scanning electron microscope (SEM, model LEO 440 VP).

## RESULTS AND DISCUSSION

The silica micro particles obtained from the experiment by different variations were analyzed using SEM

### Effect of TEOS concentration on the Silica Particle Size

The amount of TEOS in a specific reaction also plays an effective role in controlling the diameter of the silica particles. The concentration of TEOS is varied from 3.5 ml to 6.4 ml.

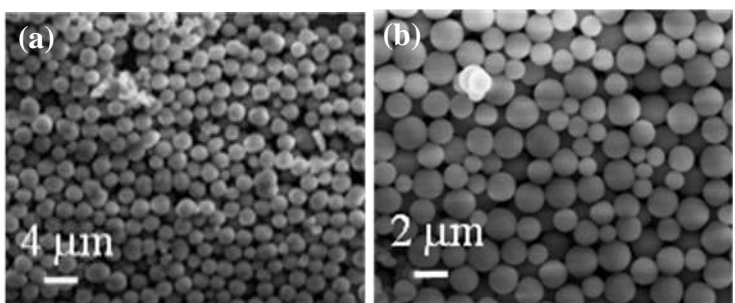

**Figure 1.** SEM images of SiO$_2$ particles synthesized using different amount of TEOS: (a) 3.5ml and (b) 6.4ml

It could be noticed from Figure 1 that there is a slight change in the size of micro particles has occurred from 2.1µm to 1.7µm as the amount of TEOS increases from 3.5ml to 6.4ml respectively by keeping all other reactants amount same. This is mainly because the initial concentration of TEOS is inversely proportional to the size of micro particles. That means higher the concentration there is smaller particles due to greater number of nucleation sites but with a greater spread of sizes.

## Effect of Ammonia concentration on Silica Particle Size

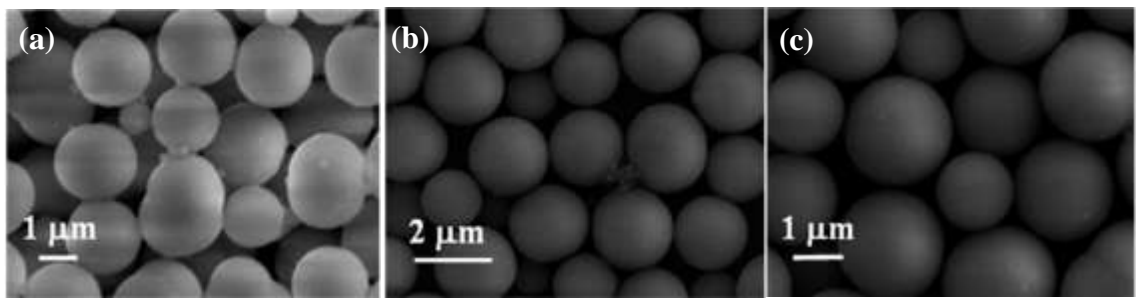

**Figure 2.** SEM images of $SiO_2$ particles synthesized using different ammonia content (a) 3ml, (b) 6ml and (c) 9ml

Figure 2 shows that with the increase in ammonia content from 3 ml to 9 ml in the solution A there is a decrease in the diameter of the micro particles from 2.1 µm to 1.7 µm. In this experiment the ammonia is mainly used as a catalyst. With the increase in ammonia concentration the quantity of the silica nuclei in the initial nucleation increases, resulting in decrease in particle size.

## CONCLUSION

Silica microspheres are synthesized using modified Stöber method. The diameter of the silica micro particles that are obtained by supplying TEOS continuously using syringe pump is much larger than traditional method. It has also been seen that the supply rate of TEOS also plays an important role in the particle size. The amount of TEOS concentration as well as ammonia concentration plays an important role in particle size control.

## ACKNOWLEDGEMENT

Authors want to thank DST-SERB Ramanujan Fellowship programme for providing financial support and also CSIR-NPL for providing instrument facility.